\documentclass[prb,american,superscriptaddress,twocolumn,showpacs,amsmath,amssymb,floatfix]{revtex4}
\usepackage[]{fontenc}
\usepackage[latin9]{inputenc}
\usepackage{array}
\usepackage{amsmath}
\usepackage{graphicx}
\usepackage{amssymb}

\usepackage{color}
\usepackage{xcolor}
\usepackage{ulem}

\makeatletter

\providecommand{\tabularnewline}{\\}

\makeatother

\usepackage{babel}

\begin{document}

\title{Charge inhomogeneities and transport in semiconductor  heterostructures with a manganese  $\delta$-layer}

\author{Vikram Tripathi} \affiliation{Department of Theoretical Physics, Tata Institute of Fundamental Research, Homi Bhabha Road, Navy Nagar, Mumbai 400005, India}
\author{Kusum Dhochak}
\affiliation{Department of Theoretical Physics, Tata Institute of Fundamental Research, Homi Bhabha Road, Navy Nagar, Mumbai 400005, India}
\author{B.A.~Aronzon} \affiliation{Russian Research Center ``Kurchatov Institute", Kurchatov Square 1, Moscow, 123182 Russia}
\affiliation{Institute for Theoretical and Applied Electrodynamics, Russian Academy of Sciences, Izhorskaya Str. 13, Moscow, 125412 Russia}
\author{V.V.~Rylkov} \affiliation{Russian Research Center ``Kurchatov Institute", Kurchatov Square 1, Moscow, 123182 Russia}
\author{A.B.~Davydov} \affiliation{Russian Research Center ``Kurchatov Institute", Kurchatov Square 1, Moscow, 123182 Russia}
\author{Bertrand Raquet} \affiliation{CNRS INSA UJF UPS, UPR 3228, Laboratoire National des Champs Magn\'{e}tiques Intenses, Universit\'{e} de Toulouse, 143 avenue de Rangueil, F-31400 Toulouse, France}
\author{Michel Goiran} \affiliation{CNRS INSA UJF UPS, UPR 3228, Laboratoire National des Champs Magn\'{e}tiques Intenses, Universit\'{e} de Toulouse, 143 avenue de Rangueil, F-31400 Toulouse, France}
\author{K.I.~Kugel} \affiliation{Institute
for Theoretical and Applied Electrodynamics, Russian Academy of
Sciences, Izhorskaya Str. 13, Moscow, 125412 Russia}

\begin{abstract}
We study experimentally and theoretically the effects of disorder,
nonlinear screening, and magnetism in semiconductor heterostructures containing a $\delta$-layer of Mn, where the charge carriers are confined within a quantum well and hence both ferromagnetism and transport are two-dimensional (2D) and differ qualitatively from their bulk counterparts. Anomalies in the electrical resistance observed in both metallic and insulating structures can be interpreted as a signature of significant ferromagnetic correlations. The insulating samples turn out to be the most interesting as they can give us valuable insights into the mechanisms of ferromagnetism in these heterostructures. At low charge carrier densities, we show how the interplay of disorder and nonlinear screening can result in the organization of the carriers in the 2D transport channel into charge droplets separated by insulating barriers. Based on such a droplet picture and including the effect of magnetic correlations, we analyze the transport properties of this set of droplets, compare it with experimental data, and find a good agreement between the model calculations and experiment. Our analysis shows that the peak or shoulder-like features observed in temperature dependence of resistance of 2D heterostructures $\delta$-doped by Mn lie significantly below the Curie temperature $T_{C}$ unlike the three-dimensional case, where it lies above and close to $T_{C}$. We also discuss the consequences of our description for understanding the mechanisms of ferromagnetism in the heterostructures under study.
\end{abstract}


\date{\today}

\maketitle

\section{Introduction}

Dilute magnetic semiconductors (DMS), incorporating semiconducting and magnetic properties within a single compound, are very promising materials for spintronics (and spintronic devices) and are also important for understanding many fundamental questions such as the origin of ferromagnetism in a semiconductor~\cite{ohno01,awschalom02,zutic04,jungwirth06,dietl07,sato10,furdina88}.
The interesting properties of DMS arise from the significant role played by magnetic exchange interactions in addition to the interactions widely studied in the conventional semiconductor structures, namely, the electron-electron Coulomb and
electron-phonon interactions, interactions with strains and random
potentials resulting from the defects and inhomogeneous distribution
of impurities. The currently most widely studied DMS materials are
those based on III-V semiconductors, in particular Mn-doped GaAs \cite{zutic04,jungwirth06,dietl07}. In such materials, if the Mn concentration is not too high, Mn substitutes Ga acting as an acceptor, so doping GaAs with Mn yields both local magnetic moments and free holes~\cite{jungwirth06,dietl07,sato10}.
One of the important lines of research here should be evidently related to the low-dimensional and, especially, two-dimensional structures given the planar character of existing microelectronic devices. In addition, heterostructures $\delta$-doped by Mn will exhibit 2D ferromagnetic behavior, which is qualitatively different from 3D ferromagnetism as there is no continuous phase transition in 2D. Nevertheless, only a limited number of studies dealing with the 2D DMS structures have been reported in the literature~\cite{awschalom04,nazmul05,wojtowicz03,aronzon10,aronzon07,vasilieva05,
wurstbauer08,rupprecht10,dietl10,dietl10a}.
In Refs.~\onlinecite{nazmul05,wojtowicz03} concerning the GaAs/AlGaAs heterostructures $\delta$-doped by Mn, the ferromagnetic (FM) state was found at rather high temperatures. However, the hole gas in these heterostructures was not quite two-dimensional since the mobility of charge carriers was so low~\cite{nazmul05} that broadening of the quantized subband levels ($\approx$300 meV) exceeded even the depth of the quantum well (150--260 meV) in this case. Such low mobility values were the result of a high density of Mn, which is responsible not only for the magnetism of the system but is also an acceptor and thus an efficient scattering center. In addition, the authors of Refs.~\onlinecite{nazmul05,wojtowicz03} aimed to provide the highest hole density just in the vicinity of Mn ions to maximize the Curie temperature $T_{C}.$

The GaAs/In$_{x}$Ga$_{1-x}$As/GaAs quantum-well structures $\delta$-doped by Mn exhibiting ferromagnetic ordering and a true 2D carrier energy spectrum were obtained by selective doping that ensured a high hole mobility (more than 2000 ${\rm cm}^{2}/{\rm V\cdot s}$ at 5 K)~\cite{aronzon10,aronzon07,vasilieva05}. The true 2D behavior in DMS heterostructures was also observed in the similar structures elsewhere; however, in these cases, the FM ordering manifested itself only at the millikelvin range of temperatures~\cite{wurstbauer08,rupprecht10,dietl10a}.

Two mechanisms underlying the FM ordering in the 2D GaAs/InGaAs/GaAs heterostructures $\delta$-doped by Mn have been proposed in the current literature~\cite{meilikhov08,menshov09}. The first model \cite{meilikhov08} attributes the FM ordering to the indirect interaction of Mn atoms by means of holes in a 2D conducting
channel. The efficiency of this mechanism is based on the large mean
free path of 2D carriers due to their remoteness from the Mn layer.
In the second model, FM ordering arises within the Mn layer, possibly mediated by the holes in the layer like in usual DMS structures \cite{menshov09}.

One of the most relevant questions is the effect of FM ordering on
the temperature dependence of resistivity, in particular, the relation of the FM ordering to the resistance anomaly (a peak or shoulder) near $T_{C}$~\cite{matsukura98}. Several theories have been proposed~\cite{moca10,timm05,matsukura04,dietl08,hwang05,majumdar98} for explaining the resistivity in bulk DMS but we are not aware of any theoretical work on transport properties of 2D DMS heterostructures. In addition to magnetism, the disorder plays a significant role in the DMS transport properties~\cite{timm03}, but again the parameters of the disorder and its effect on transport properties of the 2D DMS structures have not yet been thoroughly investigated. In particular,  a theoretical analysis providing quantitative agreement with measurements and taking into account disorder effects and peculiarities in the temperature dependence of resistivity is still required.

In this paper, we study the effect of spatial disorder of dopant concentration in the $\delta$-layer on the electronic properties of the 2D hole gas and show how at low carrier density, the competition of disorder and nonlinear screening results in the formation of ``metallic'' droplets separated by insulating regions. We make estimates for the droplet sizes and interdroplet distance, the energy level spacing in these droplet structures and the potential barrier separating neighboring droplets. Using these as parameters in a simple model for the resistivity that incorporates the effect of ferromagnetism on interdroplet tunneling, we obtain a quantitative explanation of the temperature dependence of resistivity in the DMS ferromagnetic structure with the 2D quantum well.

The rest of the paper is organized as follows. Section~ \ref{sample-setup} describes the experimental setup and the samples studied in this paper. In Section~\ref{experiments}, we discuss the available experimental evidence proving the two-dimensionality of the hole gas and the existence of ferromagnetic correlations in our samples. The model of nanoscale inhomogeneities of the hole gas is developed in Section~ \ref{droplet-model}. In Section~\ref{sec:resistivity}, we introduce a simple model for the resistivity that incorporates the effect of energy level quantization in the droplets and ferromagnetic correlation of electrons in neighboring droplets. Section~\ref{discussion} contains a discussion of our findings and their implications for the mechanisms of ferromagnetism in the DMS heterostructures.

\section{Sample and setup details}\label{sample-setup}

\begin{figure}
\includegraphics[width=0.5\columnwidth]{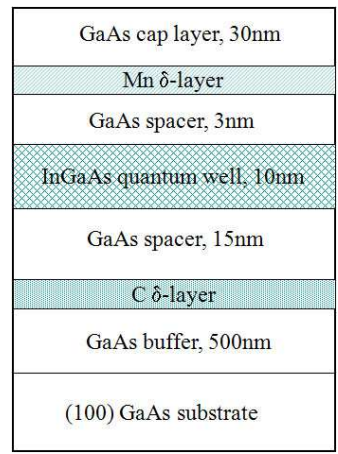} \caption{\label{fig:device} Schematic layout of the
heterostructure $\delta$-doped by Mn.}
\end{figure}

A schematic layout of the studied structures is shown in Fig.~\ref{fig:device}. %

The structure consists of an In$_{x}$Ga$_{1-x}$As quantum well (QW)
inside a GaAs matrix with a Mn $\delta$-layer separated from the
QW by a GaAs spacer of width 3 nm. The QW thickness $W$ was about 10 nm and the In fraction in it was $x\approx0.2.$ A carbon $\delta$-layer ($\approx2\times10^{12} {\rm cm}^{-2}$) was introduced  at a distance 10--15 nm below the QW just at the top of the buffer layer to compensate the hole depletion of the QW by the (undoped) buffer layer. The quantum well and the surrounding GaAs layers were grown by MOCVD at $600^{\circ}$C while the Mn $\delta$-layer and GaAs cap layers were prepared by the laser plasma deposition at $450^{\circ}$C.

The detailed description of such structures was obtained from X-ray studies reported in Ref.~\onlinecite{aronzon10}, and the methods of their growth were described in detail in Ref.~\onlinecite{vasilieva05}. The X-ray results of Ref.~\onlinecite{aronzon10} demonstrate that the Mn layer is slightly smeared forming a Ga$_{1-y}$Mn$_{y}$As region $3-5$ nm thick and with a maximum Mn content $y\leq0.05-0.08$ and not overlapping significantly with the quantum well. This is also confirmed by the values of hole mobility in the studied structures, which we find to be more than by two orders of magnitude higher than those in traditional bulk Mn-doped GaAs samples~\cite{burch06}. The mobility and other electrical and structural parameters of the studied structures are presented in Table~\ref{tab:samples}.

\begin{table*}
\begin{tabular}{|c|>{\centering}p{1.7cm}|>{\centering}p{1.1cm}|>
{\centering}p{1.5cm}|>{\centering}p{1.2cm}|>{\centering}p{1.8cm}|>{\centering}p{1.4cm}|>
{\centering}p{1.7cm}|>{\centering}p{1.8cm}|>{\centering}p{1.4cm}|>{\centering}p{1.7cm}|}
\hline
Sample  & Mn content, monolayers (cm$^{-2}\times10^{14}$)&In content $x$ & Quantum well depth $-V_0$, meV &  $V_\text{fluc}$($z$ =0, $R_c$) at 77 K, meV & Overlap probability of the hole wavefunction with Mn layer (77 K)& Hole mobility $\mu_p$ (77~K), cm$^{2}/$V$\cdot$s  & Hole density $p$ (77 K), cm$^{-2}\times10^{12}$   & Overlap probability of the hole wavefunction with Mn layer (5~K) & Hole mobility $\mu_p$ (5 K), cm$^{2}/$V$\cdot$s  & Hole density $p$ (5~K), cm$^{-2}\times10^{12}$ \tabularnewline
\hline
1 & 1.2 (6.0)& 0.18  & 85 & 260 & 0.15$\times10^{-2}$  & 1350 & 1.8  & 0.15$\times10^{-3}$ & 180  & 0.3 \tabularnewline
\hline
2 & 0.5 (3.0)& 0.21 & 100 & 170 & 0.51$\times10^{-2}$  & 1860 & 2.0  & 0.52$\times10^{-3}$ & 2950  & 0.71 \tabularnewline
\hline
3 & 0.4 (2.5)& 0.23 & 115 & 160& 0.39$\times10^{-2}$   &1930 & 1.8  & 0.63$\times10^{-3}$ & 3240  & 0.79 \tabularnewline
\hline
4 & 0.35 (2.0)& 0.17  & 70 & 145& 0.72$\times10^{-2}$   & 2370 & 1.4  & 0.9$\times10^{-3}$ & 3400  & 0.46\tabularnewline
\hline
5 & 0 & 0.18 & 85 &-- & -- & 1600   & 0.5  & --& --  & --\tabularnewline
\hline
\end{tabular}
\caption{\label{tab:samples} Parameters characterizing the samples under study. Samples 1--4 are $\delta$-doped by Mn. Sample 5 is $\delta$-doped by carbon instead of Mn. All the samples have a carbon layer too as shown in Fig.~\ref{fig:device}. We also present the  model estimates for the fluctuation potential $V_{\text{fluc}}$ at the quantum well edge facing the Mn dopant layer, and the overlap probability of the hole wavefunction with a 1 nm thick region centered at the $\delta$-layer of Mn situated 3 nm away from the quantum well. Here, $R_c$ is the screening length. At each In content, the quantum well depth was estimated using the known experimental results, according to which the valence band discontinuity is about 1/3  of the band gap discontinuity~\cite{arent89}.}
\end{table*}

As we have already mentioned, the main feature of these structures is that they are really two-dimensional and exhibit FM ordering at relatively high  temperatures~\cite{aronzon10,aronzon08}. The two-dimensionality is confirmed by our observation of Shubnikov-de Haas oscillations  and of the quantum Hall effect. We expect that the transport is mostly due to light holes because of the large splitting between the light and heavy hole $\Gamma_8$ subbands (about 90 meV for In$_{x}$Ga$_{1-x}$As for $x\approx0.2$)~\cite{schirber05,grisha05}. This splitting arises from the biaxial strain caused by the lattice mismatch of GaAs and In$_{x}$Ga$_{1-x}$As and results in light mass behavior of holes (see Ref.~\onlinecite{drachenko09} and references therein).

\begin{figure}
\includegraphics[width=0.95\columnwidth]{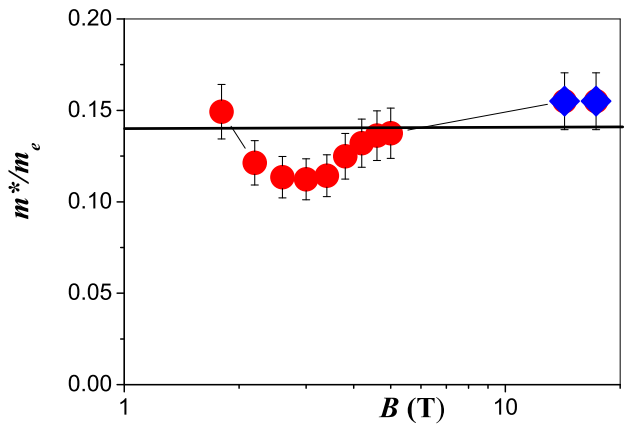}
\caption{\label{fig:eff-mass}(color online) Effective mass dependence on magnetic field for Sample 2 (red circles) measured by ShdH oscillations and for Sample 4 (blue diamonds) measured by cyclotron resonance. Solid black line corresponds to $m^*/m_{e}=0.14$.
}
\end{figure}

To find the actual value of the effective mass in our structures, which is important for calculations of the energy levels in quantum wells and for the adequate interpretation of the transport data, we performed  cyclotron resonance and Shubnikov--de Haas (ShdH) oscillations measurements (see Fig.~\ref{fig:eff-mass}).  The cyclotron resonance measurements were performed at the Toulouse High Magnetic Field Laboratory (LNCMP) using a long-pulse coil, delivering magnetic fields up to 40 T with a total pulse length of 800 ms. The obtained values of the hole effective mass $m^* \simeq 0.14 m_{e}$ are in agreement with $m^*$ determined in nonmagnetic GaAs/InGaAs/GaAs heterostructures from the ShdH oscillations~\cite{schirber05} as well as from the recent cyclotron resonance measurements~\cite{drachenko09}.

In our opinion, such samples are optimal for studies of disorder and
magnetic properties in a two-dimensional hole gas. We have performed
measurements using samples with different Mn content to understand
the effect of disorder and doping on the crossover from the metallic-like to insulating behavior. This crossover is also affected by the degree of ionization of the Mn dopants as we will see below. Samples for transport measurements were prepared by photolithography and have a Hall-bar geometry of width 0.3 mm between the Hall probes and 1.5 mm between resistance probes. Measurements of the temperature and magnetic field dependence of sample resistance and Hall effect were performed in the $5-300$ K temperature range at magnetic fields up to 3 T.

\section{Experimental observations and their consequences} \label{experiments}

The main purpose of this paper is the quantitative description of
disorder and peculiarities of temperature dependence of the resistance $R(T)$ in 2D DMS structures. In this part of the paper, we summarize and present experimental results needed for this quantitative explanation. For that, we need first, to recall that the metal to insulator transition occurs in dilute magnetic semiconductors with increase of Mn content, to present the data confirming that it indeed takes place in 2D heterostructures, such as our samples, and to provide arguments for its percolative nature; second, to illustrate the 2D character of the electron energy spectrum in our case; third, to provide evidence for FM correlations; and fourth, to present experimental data on peculiarities of $R(T)$ related to FM ordering. Here we will present experimental data in accordance with each of these items.

\begin{figure}
\includegraphics[width=0.95\columnwidth]{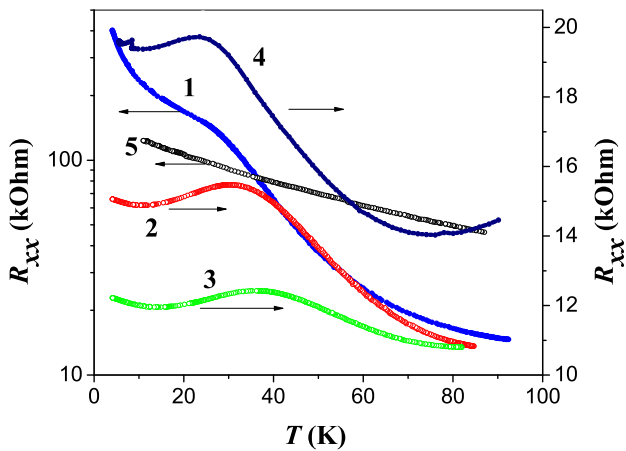} \caption{\label{fig:resistance-data}(color online) Resistance data for the Mn $\delta$-doped heterostructures (1, 2, 3, and 4) for
different carrier and doping densities (see Table~\ref{tab:samples})
and a carbon $\delta$-doped heterostructure (5). Note the absence
of any resistance anomaly in the carbon $\delta$-doped sample, while
the Mn $\delta$-doped samples exhibit an anomaly (hump or shoulder),
which is likely due to the magnetic ordering.}
\end{figure}

1. The temperature dependence of resistance $R(T)$ for all samples
listed in Table~\ref{tab:samples} is plotted in Fig.~\ref{fig:resistance-data} (some of these data were also presented in Ref.~\onlinecite{aronzon10}).

It is seen that the low-temperature resistance of these samples ranges from 10 kOhm to 500 kOhm and the ratio $R(5K)/R(70K)$ also drastically changes for samples with different Mn content (about 30 for Sample 1 and 1.1(5) for Sample 3). This suggests that we have a set of samples ranging from very insulating to nearly metallic. Note that strictly speaking even the most ``metallic'' samples are not the classical metals since their resistance, although rather low, increases with lowering temperature in the range  40--100 K. Samples 2 and 3 also show a larger mobility at 5 K compared to 77 K, which is indicative of metallic behavior while Sample 1 has lower mobility at 5 K, as should be the case with an insulator. Note also  that the resistivity of our samples turns out to be of the same order of magnitude as the value $\rho \sim 0.2h/e^2$, at which the metal-insulator transition occurs in 2D DMS structures~\cite{Jar07}.

\begin{figure}
\includegraphics[width=0.95\columnwidth]{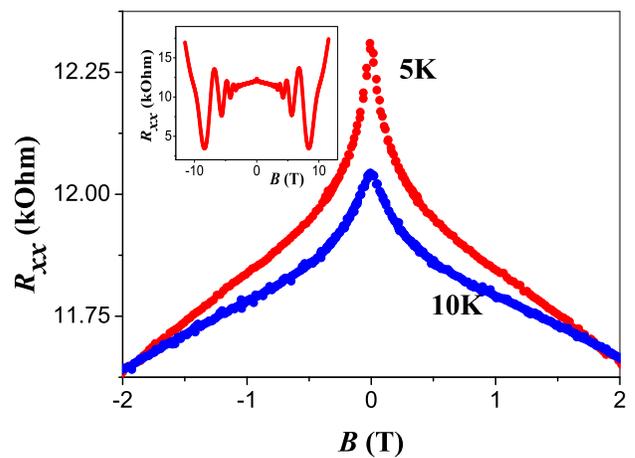} \caption{\label{fig:magnetoresistance}(color online) Magnetic field dependence of the resistance of Sample 3 at different temperatures. The inset shows Shubnikov--de Haas (ShdH) oscillations indicating metallicity. }
\end{figure}

The existence of well pronounced Shubnikov--de Haas (ShdH) oscillations in Samples 2 and 3 tells us that these two samples are on the metallic side of the percolation transition. In Fig.~\ref{fig:magnetoresistance}, we show the magnetic field dependence of the resistance at two values of the temperature. The inset shows ShdH oscillations previously discussed in Ref.~\onlinecite{aronzon10}. In general, the magnetic field dependence of the resistance is determined by both the proximity to the ferromagnetic transition and quantum corrections. Our low-temperature ($T\ll T_{C}$) negative magnetoresistance occurs evidently not due to any spin phenomenon but is related to the destructive effect of the magnetic field on quantum corrections to the sample conductivity related to interference of scattered carriers as was pointed out for bulk DMS materials in Ref.~\onlinecite{timm05}. In fact, in the range 0.04--0.3 T, the observed conductivity is proportional to $\log(B)$ as it should be for weak localization corrections in 2D. In contrast to the results of Ref.~\onlinecite{grisha05},  where quantum corrections to conductivity were studied in a similar structure doped by C instead of Mn, we did not observe antilocalization. In our case, the absence of antilocalization is due to the hole spin splitting caused by ferromagnetic ordering. The negative magnetoresistance presented in Fig.~\ref{fig:magnetoresistance} resulting from weak localization is a signature of the important role of disorder in transport properties of the studied samples.

\begin{figure}
\includegraphics[width=0.95\columnwidth]{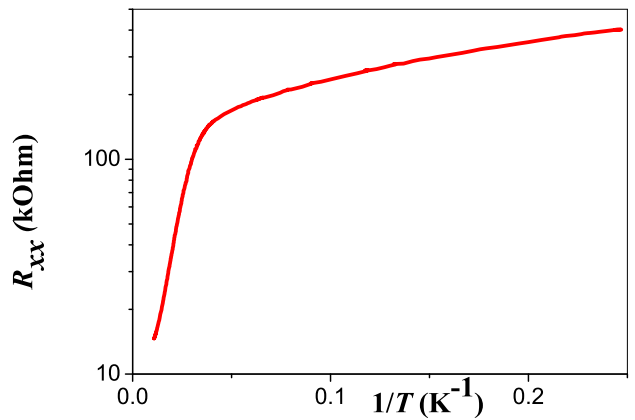} \caption{\label{fig:sample-4834}(color online) Plot of $\log R(T)$ vs $T$ for Sample 1 demonstrating an Arrhenius behavior for temperatures higher than about 30 K and a hopping behavior at low temperatures.}
\end{figure}

In contrast to Samples 2 and 3, Sample 1 is quite insulating, with
$R(5K)/R(70K)\approx 30.$ The resistance $R(T)$ exhibits an Arrhenius behavior (activation energy $\approx110$ K for $T\gtrsim30$ K) with crossover to the hopping regime at temperatures less than 30 K (see Fig.~\ref{fig:sample-4834}).

Sample 4 with a smaller carrier density compared to Samples 2 and 3 is closer to the percolation transition having high enough
values of $R(5K)=19.7$ kOhm and $R(5K)/R(70K)\approx 1.5.$
We found that the temperature dependence of the resistivity can be fitted to the Arrhenius law (activation energy $\approx20$ K for $T\gtrsim30$ K).

Thus, we have a wide enough set of samples on both sides of the metal-insulator transition. We believe that this transition is of the percolation type because, as we find below, the parameter $k_{F}l$ near the percolation threshold is greater than unity ($l$ is the hole mean free path or scattering length). For the above mentioned set of samples, we will calculate in theoretical part parameters of fluctuation potential (or disorder) and the electronic structure in the quantum well, which consists of metallic droplets separated by insulating barriers. The insulating samples are most interesting as they can give us valuable insights into the mechanism of ferromagnetism in these DMS heterostructures and we will concentrate on these samples. We believe all the studied samples are close to the percolation transition since all of them demonstrate some features of both metallic and insulating behavior.

2. For comparison with the theoretical calculations, we need to prove that the hole gas in our samples has a 2D energy spectrum.

The 2D character of Samples 2 and 3 is proved by ShdH oscillations, which are observed only when the magnetic field is perpendicular to the sample plane as seen in the inset of Fig.~\ref{fig:magnetoresistance}. Manifestations of the quantum Hall effect (QHE) were observed even in the most insulating Sample 1 (see Ref.~\onlinecite{aronzon10}), which establishes its 2D nature. The existence of the QHE on the insulating side of the percolation transition can be explained following the arguments presented in Refs.~\onlinecite{efros88}.
Near the percolation transition, a sample consists of ``metallic'' droplets and insulating regions, and in 2D, the Hall constant $R(x_m)$ is related to the conductivity $\sigma(x_m)$~\cite{shklovskii77,aronzon10} as
\begin{align}
R(x_m) & \approx R_{m}\left[1-\frac{\sigma_{d}^2}
{\sigma^{2}(x_m)}\right],\label{eq:hall}\end{align}
where $x_m$ is the fraction of the metallic phase, and the
subscripts $d$ and $m$ refer to the insulating and metallic regions.
Under conditions of finite tunneling between metallic droplets,
the Hall constant is approximately equal to $R_{m}$ in some region near the percolation transition even when the sample is in the insulating phase~\cite{aronzon10}. Experimental observation of QHE in disordered insulators has also been reported elsewhere~\cite{hilke98,ilani04}.

3. The evidence for ferromagnetic (FM) correlations comes from the observation of a hump or shoulder in the temperature dependence of resistivity as presented in Fig.~\ref{fig:resistance-data}. The fact that this feature is observed for all samples doped by Mn but is absent for Sample 5 doped by C instead of Mn shows that it has a magnetic origin. The direct evidence of FM ordering for Samples 1 and 4 was through the observation of a hysteresis loop in the magnetization curve~\cite{aronzon07,aronzon08-2,aronzon08-3}. The observation of anomalous Hall effect (AHE) in all mentioned samples ~\cite{aronzon10,aronzon08,aronzon07} is yet another evidence.

4. The temperature dependence of resistivity $R(T)$ is presented
in Fig.~\ref{fig:resistance-data}. It is commonly accepted that the
``anomalous'' hump or shoulder of this temperature dependent resistance could be used as a measure of the Curie temperature \cite{jungwirth06,matsukura98,matsukura04}. There are differing opinions on whether the anomaly in $R(T)$ or $dR/dT$ should be accepted as $T_{C}$ \cite{novak08}. While it is justified for the case of bulk metals to associate the temperature, at which the anomaly occurs, with $T_{C},$ we will show below that
in the 2D case, the situation is quite different. A comparison of our experimental results and theoretical calculations of $R(T)$ for
Samples 1 and 4 is presented below in Fig.~\ref{fig:fits}.

Thus, now we have a good basis to formulate a theoretical model for
the charge distribution in 2D $\delta$-doped DMS heterostructures,
which can be used further on for calculations of the temperature dependence of resistivity.

\section{Model of nanoscale inhomogeneities}\label{droplet-model}

For the purpose of analysis, we can consider the following system,
which captures the main physics. The two-dimensional hole gas (2DHG)
is formed within the InGaAs quantum well. The holes
in the 2DHG are provided by Mn acceptors distributed in a $\delta$-layer with density $n_{a}$, which is spatially separated from the quantum well by a GaAs spacer with thickness $\lambda.$ We thus have two interacting subsystems - the $\delta$-layer, where the Mn atoms are a source of holes as well as of magnetism owing to their spins, and the quantum well, where the behavior of the holes is affected by the charge and spin of Mn atoms. In addition, the holes in the 2DHG are known to affect the distribution of magnetization in the $\delta$-layer. This will be particularly true for the more metallic samples~\cite{meilikhov08}.

The parameters characterizing the samples under study are listed in
Table~\ref{tab:samples}. The table shows the total Mn content in
the $\delta$-layer, the quantum well depth $V_0$ in the absence of fluctuations, the hole densities $p$, and mobilities $\mu_{p}$ in the 2DHG layer at two different temperatures on either side of the ferromagnetic transition.

At low carrier density, it has been shown \cite{gergel78,pikus89,nixon90,fogler04,tripathi06}
that the interplay of disorder (due to random potentials of the charged Mn atoms) and nonlinear screening by the holes can lead to inhomogeneities in the carrier density. The physical picture of droplet formation and metal-insulator transition is as follows. Charge fluctuations of the ionized Mn acceptors create a fluctuating potential for the hole gas in the quantum well. The holes begin  filling the deepest energy levels in the potential relief. Introduction of holes also affects the size of the potential fluctuation because of screening. We assume a Gaussian white noise distribution for the charge density $\rho(\mathbf{r},z)=en(\mathbf{r})\delta(z+\lambda)$ of the Mn atoms in the $\delta$-layer ($z$ axis is directed perpendicular to the $\delta$-layer, $z$ = 0 corresponds to the GaAs/InGaAs interface). For points $\mathbf{r},$ $\mathbf{r}'$ lying in the $\delta$-layer we have,
\begin{align}
\langle n(\mathbf{r})n(\mathbf{r}')\rangle-\langle n^{2}\rangle & =n_{a}'\delta(\mathbf{r}-\mathbf{r}'),\label{eq:gaussian}\end{align}
where $n_{a}'$ is the total density of negative ionized acceptors  and positively charged compensating donors in the Mn $\delta$-layer: $n_{a}'  =  n_{a}^- + n_{d}^+$. In actual heterostructures, the ionization is usually partial due to several causes: (a) Mn atoms could substitute Ga being acceptors or enter interstitial positions acting as donors, thus leading to a compensation. Comparing Mn content in the Mn doped layer, which is in fact Ga$_{1-y}$Mn$_y$As, with results for bulk material it is natural to suggest that percentage of Mn in interstitial position is about 10\%, such results are summarized in review articles Refs.~\onlinecite{jungwirth06, sato10}; (b) Mn atoms could form compounds with Ga and As;  (c) there also exists the density correlation of the dopants related to their frozen nonequilibrium distribution~\cite{pikus89}. According to the latter effect, for example, for Sample 4, we have estimated $n_{a}'\sim0.06n_{a}$, mimicking an ionization degree of 0.06. In the further calculations, for simplicity and consistency, we take a typical value of $n_{a}'$ and hence assume that $n_{a}'= 0.1n_{a}$, which is in agreement with the effective ionization of about 0.1 observed in Ga$_{1-y}$Mn$_y$As samples~\cite{jungwirth06,sato10} although the actual degree of ionization degree can be even smaller.

From Eq. \eqref{eq:gaussian},  it is easy to see that the variance of the fluctuation charge density in a circular region of size $R$ is $\langle\delta n^{2}(R)\rangle=n_{a}'/(\pi R^{2}).$
The random distribution of charges creates a fluctuating potential
$\phi$ at the interface. In the presence of holes in the 2DHG, the
potential fluctuations are screened beyond a length scale $R_{c}$
where the fluctuation charge density
$\sqrt{\langle\delta n^{2}(R_{c})\rangle}=\sqrt{n_{a}'/\pi}/R_{c}$
becomes less than the hole density $p.$ The variance of the
potential fluctuations at the interface is \cite{tripathi06}
\begin{align} \langle\delta\phi^{2}\rangle  & =\frac{n_{a}'e^{2}}{8\pi\kappa^{2}\epsilon_{0}^{2}}\left\{ \ln\left[\frac{4d^{2}}{\lambda(2d-\lambda)}\right]
\right.\nonumber \\
  &\!\!\!\!\!\!\!\left.-2\ln\left[\left(\frac{\lambda^{2}+R_{c}^{2}}{(2d-\lambda)^{2}
+R_{c}^{2}}\right)^{\frac{1}{4}}+\left(\frac{(2d-\lambda)^{2}+R_{c}^{2}}
{\lambda^{2}+R_{c}^{2}}\right)^{\frac{1}{4}}\right]\right\} .\label{eq:pot-fluc-exact}\end{align}
Here, $\kappa=12.9$ is the permittivity of GaAs and $R_{c}={n_{a}'}^{1/2}/\pi p$ is the characteristic screening length described above. Parameter $d$ is a length scale, beyond which the potential fluctuations get screened even in the absence of holes in the quantum well. Often there is a metallic gate on the sample, in which case $d$ is equal to the distance from the quantum well to the gate.

In cases where the inequality $2d\gg R_{c},\lambda$ is met, the potential fluctuations can be expressed in a much simpler form,
\begin{align}
\langle\delta\phi^{2}\rangle & \approx\frac{n_{a}'e^{2}}{16\pi\kappa^{2}\epsilon_{0}^{2}}
\ln\left[1+\left(\frac{R_{c}}{\lambda}\right)^{2}\right].
\label{eq:pot-fluc}
\end{align}

The holes in the quantum well are centered at a distance $z_{0}$
(measured from the interface closest to the Mn layer) in the
direction perpendicular to the interface. To obtain $z_{0},$
we solve the Schr\"{o}dinger equation in the quantum well in
the quantum well taking into account the ($z$-dependent)fluctuating potential,

\begin{align}
\left[-\frac{\hbar^2}{2 m^{*}}\frac{d^2}{d z^2}+V(z)\right]\psi_{n}=E_{n}\psi_{n}.
\label{eq:QW}
\end{align}

\begin{figure}
\includegraphics[width=0.95\columnwidth]{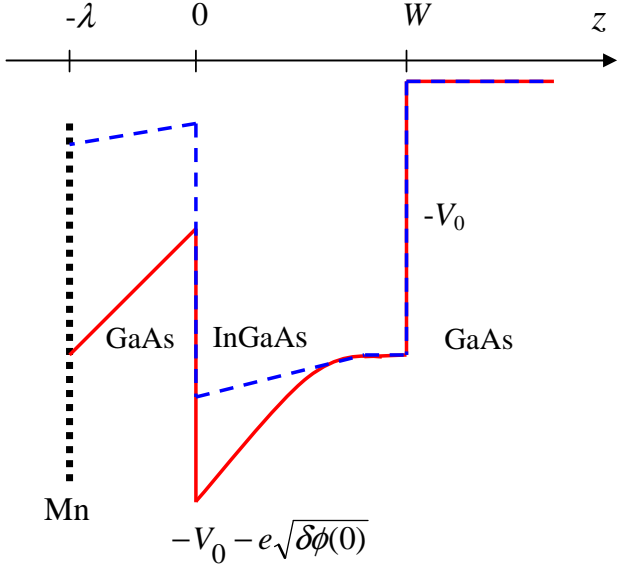} \caption{\label{fig:Well}(color online) Schematic of the quantum well potential (shown inverted). Dashed (blue) line represents the quantum well potential in the absence of fluctuations and the solid (red) line shows the potential well with
an attractive fluctuation potential. The dotted line indicates the Mn dopants at a distance $\lambda$ from the left face of the quantum well. The quantum well of thickness $W$ is defined in the InGaAs layer sandwiched between GaAs regions.}
\end{figure}

Here $V(z)$ is the quantum well potential together with the
fluctuations (see Fig.~\ref{fig:Well}). $n=1,2,3\ldots$ refers to
the subband index> For holes, we use the approximation of the parabolic dispersion with the effective mass $m^{*}=0.14m_{e}$  as measured from cyclotron resonance for these structures.
We approximate $V(z)$ as follows. For $z<0,$
$V(z) = \alpha(|z+\lambda|-\lambda),$
$\alpha = ep/\kappa\epsilon_0 .$ For $z>W,$ we have $V(z)=0$,
where $W$ is the quantum well thickness. For $0<z<W,$  we have
$V(z)=V_{QW}(z)-e\sqrt{\langle\delta\phi^2(R_c,z)\rangle}$,
where $V_{QW}(z)=-V_{0}+\alpha z.$ For the present devices,
we have taken $W=10$ nm, $\lambda=3$ nm and the values of
$V_0$ are as shown in Table~\ref{tab:samples}. We also assume
that the spatially varying fluctuation potential does not affect
the valence band position away from the quantum well.
The condition for existence of a subband is $E_{n}<0.$
Table~\ref{tab:samples} shows the overlap probability
$\int_{\delta z} dz |\psi_{1}(z)|^2,$ where $\delta z$ is a
1 nm thick region centered at the $\delta$-layer.

We also find the hole wavefunction in the GaAs region
decreases away from the quantum well with a localization length
$\xi_z$ having a value ranging from 1 nm to 2 nm. This is
comparable to the localization length estimated in
Ref.~\onlinecite{meilikhov08}. We will henceforth use
$\lambda + z_0$ as the distance of the hole gas from
the $\delta$-layer for the purpose of calculating the
potential fluctuations. For a given subband $n,$ we determine
$z_{0,n}$ as $z_{0,n}=\int dz\,z|\psi_{n}(z)|^2.$
Tables~\ref{tab:droplets77K} and \ref{tab:droplets5K}
represent the values of $E_{n}$ and $z_{0,n}$ for the first
two subbands. For simplicity, we will from now on denote
$e\sqrt{\langle\delta\phi^{2}(d+z_{0,n},\lambda+z_{0,n},R)\rangle}$
by $V_{\text{fluc}}(z_{0,n},R).$ The values of the fluctuation
potential are also shown in Table~\ref{tab:samples}.

Now, we describe how at low enough density, the holes in the 2DHG can get organized into charge droplets. Let $R_{p,n}$ be the size of a droplet. The potential fluctuations associated with this length scale are given by$V_{\text{fluc}}(z_{0,n},R_{p,n}).$ Suppose that the holes fill this potential well up to a wavevector $k_{\text{max}}.$ From the virial theorem,
\begin{align}
\frac{\hbar^{2}k_{\text{max},n}^{2}}{2m^{*}} & =\frac{1}{2}V_{\text{fluc}}(z_{0,n},R_{p,n}),\label{eq:virial1}\end{align}
where the factor of $1/2$ is for a linear-in-$R_{p,n}$ confining
potential, which is approximately the case here. The number of occupied  states in the droplet is approximately $(k_{\text{max}}R_{p})^{2}/2,$ which can be equated with the fluctuation charge $N_{h}=\pi R_{p}^{2}\times\sqrt{n_{a}'/\pi}/R_{p}=
\sqrt{\pi n_{a}'}R_{p}$, if only the lowest subband is occupied. We will discuss below the case, where more than one subband is occupied. If only the lowest subband is occupied and $2d\gg R_{c},\lambda$ , Eq.~\eqref{eq:virial1} yields a very simple solution for the droplet size
\begin{align}
R_{p,1} & \approx\sqrt{2a_{B}(\lambda+z_{0,1})}.\label{eq:puddle-size}\end{align}
 Eq.~\eqref{eq:puddle-size} is valid when $\lambda+z_{0}$ is much
greater than $a_{B}.$ The Bohr radius corresponding to these parameters is $a_{B}\approx 5.3$ nm.  In our case $\lambda+z_{0} \approx a_B,$ and this approximation does not give the correct values. Therefore we solve for $R_p$ numerically using Eqs.~\eqref{eq:virial1} and \eqref{eq:Emax1}. The droplet size $R_{p}$ and the number of holes per droplet are only weakly dependent on $p.$

Now, let us discuss the case where two subbands are occupied. The energy of the highest occupied state measured from the bottom of the lowest subbands ($n=1$) is of the order of (we will obtain a better estimate below)
\begin{align}
E_{\text{max},1} & =\frac{\hbar^{2}k_{\text{max},1}}{2m^{*}}=\frac{\hbar^{2}\sqrt{\pi n_{a}'}}{m^{*}R_{p,1}}.\label{eq:Emax1}\end{align}

\begin{figure}
\includegraphics[width=0.95\columnwidth]{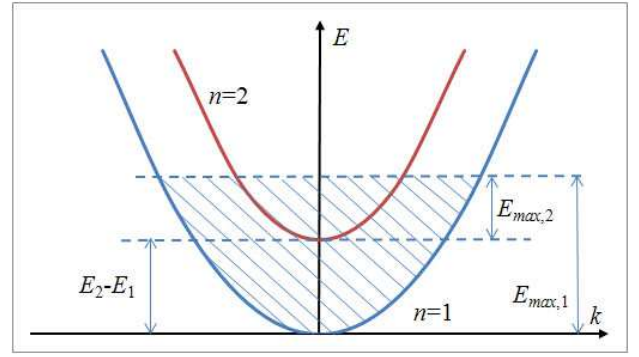}
\caption{\label{fig:2-bands} A schematic picture of the dispersion curves $E(k)$ (at small $k$) corresponding to the two lowest subbands $n=1$ and $n=2$. $E_1$ and $E_2$ are the fluctuation potentials corresponding to $z_{0,1}$ and $z_{0,2}$ respectively, and $R=R_c$. The shaded region represents filled states. The two subbands have a common chemical potential.}
\end{figure}

From this estimate of $E_{\text{max},1}$ and the energies $E_{1}$
and $E_{2},$ we can see that the second subband is also partially
occupied for Sample 1. It may seem natural to estimate the droplet sizes $R_{p,n}$ of the two subbands independently, in which case we would get the droplet sizes $R_{p,n}\approx\sqrt{2a_{B}(\lambda+z_{0,n})}.$ However, the
filling of the two subbands is not independent and the following
two conditions need to be satisfied in addition to the relations in
Eq.~\eqref{eq:virial1}. First, the chemical potential of the droplets corresponding to the two subbands should be the same (see Fig.~\ref{fig:2-bands})
\begin{align}
E_{\text{max},1}-E_{\text{max},2} & =\frac{\hbar^{2}}{2m^{*}}(k_{\text{max},1}^{2}-k_{\text{max},2}^{2})=
E_{2}-E_{1}.\label{eq:common-E}\end{align}

Second, the total number of bound holes is now distributed over the two bands. This effectively results in the transfer of some of the higher energy holes in the lower subband to lower energy empty states in the upper subband. This would lead to a decrease of the droplet
size corresponding to the lower subband, and a finite droplet size
in the upper subband. The transfer of charge from the lower subband
to the upper subband naturally makes higher the concentration of charge in the droplet since the charge carriers can occupy two bands in the same region.

We must now satisfy the following relationship
\begin{align}
\sqrt{\pi n_{a}'}R_{p,1} & =\frac{(k_{\text{max},1}R_{p,1})^{2}}{2}+
\frac{(k_{\text{max},2}R_{p,2})^{2}}{2}.
\label{eq:charge-cons}\end{align}
 Eqs. \eqref{eq:common-E}, \eqref{eq:charge-cons}, and \eqref{eq:virial1} form a system of coupled nonlinear equations that may be solved for $E_{\text{max},n}$ and $R_{p,n}.$ Tables~\ref{tab:droplets77K} and \ref{tab:droplets5K} show the calculated values of $E_{\text{max},1}$, $R_{p,1}$, and $R_{p,2}$.

Now we analyze the conditions for a metal-insulator crossover. The
localization length, which characterizes the spread of the hole wavefunction outside the droplets is
\begin{align}
\xi & =\frac{\hbar}{\sqrt{2m^{*}(|E_1|-|V_{QW}(z_0)|-
|E_{\text{max},1}|)}}.\label{eq:xi-loc}\end{align}
A percolation transition to a more conducting regime is expected
when the droplets begin to overlap. The droplets are said to ``overlap'' once the interdroplet tunneling becomes significant; in other words, the localization length $\xi$ of holes in the droplets becomes comparable to the separation $D_{1}$ between the surfaces of neighboring droplets ($D_{1}/\xi\sim1$). To obtain the separation of the droplets, we note that the total number of holes in a droplet with both bands considered is $N_{h}=\sqrt{\pi n_{a}'}R_{p,1}.$ These holes are ``drained'' from an area of size $R$ such that $N_{h}=\pi R^{2}p.$ We thus get the size of the catchment area of a droplet, $R=\sqrt{R_{p,1}R_{c}}$. The distance $D_{1}$ between the surfaces of neighboring droplets is then $D_{1}=2(\sqrt{R_{p,1}R_{c}}-R_{p,1}).$ Assuming that the potential well corresponding to the second subband is also centered at the well corresponding to the first subband, we find the distance between the droplets corresponding specifically to holes in the second subband, $D_{2}=2(\sqrt{R_{p,1}R_{c}}-R_{p,2})$.
\begin{table*}
\begin{tabular}{|>{\centering}p{1.3cm}|>{\centering}p{1.3cm}|>
{\centering}p{1.3cm}|>{\centering}p{1.3cm}|>{\centering}p{1.3cm}|>
{\centering}p{1.3cm}|>{\centering}p{1.3cm}|>{\centering}p{1.3cm}|>
{\centering}p{1.8cm}|>{\centering}p{1.3cm}|>{\centering}p{1.3cm}|>
{\centering}p{1.3cm}|}
\hline
Sample  & $R_{c}$ (nm)  & $z_{0,1}$(nm)  & $E_{1}$(meV)  & $z_{0,2}$(nm)  & $E_{2}$(meV)  & $R_{p,1}$(nm)  & $R_{p,2}$(nm)  & $E_{\text{max},1}$(meV)  & $D_{1}$(nm)  & $D_{2}$(nm)  & $\xi$(nm)\tabularnewline
\hline
\hline
1  & 24.28  & 1.79  & -203 & 3.62  & -80 & 8.96 & 0 & 83  & 11.58  & --  & 1.82 \tabularnewline
\hline
2  & 15.45  & 1.57  & -142  & 0.78  & -33  & 8.79  & 0  & 60 & 5.72 & --  & 3.50 \tabularnewline
\hline
3  & 15.67  & 1.71  & -147  & 2.30  & -40 & 8.90 & 0  & 54  & 5.82 & --  & 3.70 \tabularnewline
\hline
4  & 18.02  & 1.71  & -106  & 1.34  & -17 & 8.90  & 0  & 48  & 7.53  & --  & 3.46 \tabularnewline
\hline
\end{tabular}

\caption{\label{tab:droplets77K} Calculated values at a temperature of 77 K for the screening length $R_{c},$ droplet sizes $R_{p,n},$ droplet separations $D_{n}$ corresponding to $R_{p,n},$ penetration depths $z_{0,n},$ energy levels $E_{n},$ the maximum energy, $E_{\text{max},1},$ of occupied states measured from the bottom of the potential well for the lowest subbands ($n=1$) and the localization length $\xi.$ The calculations are for an effective $n_{a}'=0.1n_{a}.$ Note that in the last three samples, the separation of the droplets is comparable with the localization length, implying proximity to the ``metallic'' phase.}
\end{table*}

\begin{table*}
\begin{tabular}{|>{\centering}p{1.3cm}|>{\centering}p{1.3cm}|>
{\centering}p{1.3cm}|>{\centering}p{1.3cm}|>{\centering}p{1.3cm}|>
{\centering}p{1.3cm}|>{\centering}p{1.3cm}|>{\centering}p{1.3cm}|>
{\centering}p{1.8cm}|>{\centering}p{1.3cm}|>{\centering}p{1.3cm}|>
{\centering}p{1.3cm}|}
\hline
Sample  & $R_{c}$ (nm)  & $z_{0,1}$(nm)  & $E_{1}$(meV)  & $z_{0,2}$(nm)  & $E_{2}$(meV)  & $R_{p,1}$(nm)  & $R_{p,2}$(nm)  & $E_{\text{max},1}$(meV)  & $D_{1}$(nm)  & $D_{2}$(nm)  & $\xi$(nm)\tabularnewline
\hline
\hline
1  & 145.7  & 2.99  & -275  & 5.37  & -204  & 9.8  & 0.6 & 76  & 55.59  & 73.59  & 1.45 \tabularnewline
\hline
2  & 43.52  & 2.64  & -206  & 5.16  & -128  & 9.57  & 0  & 55 & 21.68 & --  & 1.93 \tabularnewline
\hline
3  & 35.71  & 2.58  & -198  & 5.10  & -119  & 9.51  & 0  & 51  & 17.83  & --  & 2.13 \tabularnewline
\hline
4  & 54.85  & 2.93  & -155  & 5.36  & -91  & 9.77  & 0  & 41  & 26.75 & --  & 2.11\tabularnewline
\hline
\end{tabular}

\caption{\label{tab:droplets5K} Calculated values at a temperature of 5 K for the same quantities as described in Table~\ref{tab:droplets77K} for an effective $n_{a}'=0.1n_{a}.$ Note that all the samples are
found to be well-insulating at this temperature. The interdroplet
distance $D_{1}$ is very sensitive to the hole density $p,$
and since $p$ increases rapidly with temperature, the ratio $D_{1}/\xi$ can become comparable to unity at relatively low temperatures enabling a transition to the ``metallic'' phase.}
\end{table*}

From Tables~\ref{tab:droplets77K} and \ref{tab:droplets5K}, it is
clear that the droplet size is fairly constant for different temperatures and hole densities. Sample 1 is insulating at all temperatures. The behavior of the remaining samples differs significantly for $T=77$ K and $T=5$ K. For these samples at $77$ K, the interdroplet separation is comparable to the localization length $\xi$ which means that they are more ``metallic''. Note that the interdroplet distance is larger for Sample 4, which gives rise to an Arrhenius-type behavior in a wide enough temperature range. At $T=5$ K, the interdroplet separation far exceeds $\xi$ so that all the samples are in the insulating regime. That does not agree with experimental results because Samples 2 and 3 exhibit a quasimetallic behavior even at $T=5$ K. This could result from the strong dependence of $D_{1}$ on the sample parameters (the carrier density, for example) at low temperatures. Thus, the droplet picture  following from our calculations being quite reasonable at $T=77$ K, may give overestimated values of the interdroplet distances at $T=5$ K. Since $D_{1}$ is very sensitive to the carrier density $p,$ and $p$ changes rapidly with temperature, the insulator to metal crossover will take place in Samples 2-4 as the temperature is increased from 5 K. We also find that in contrast to the usual situation encountered in GaAs heterostructures, where the contribution of all but the lowest subbands can be neglected, in Sample 1, the second subband is also occupied.

\begin{figure}
\includegraphics[width=0.95\columnwidth]{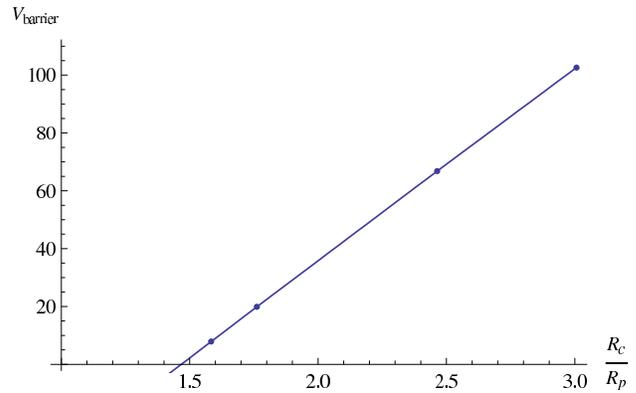}
\caption{\label{fig:barrier} Plot of the potential barrier (for Sample 3) $V_{\text{barrier}}$ for holes at the Fermi levels in the droplets as a function of the parameter $R_{c}/R_{p}$.}
\end{figure}

The energy level spacing, $\Delta,$ of a droplet can be found by
noting that addition of a hole to a droplet increases $R_{p,1}$ by
an amount $1/\sqrt{\pi n_{a}'}.$ The difference of the values of $E_{\text{max},1}$ of the droplets of size $R_{p,1}+1/\sqrt{\pi n_{a}'}$ and $R_{p,1}$ respectively gives us the level spacing at the chemical potential. The level spacing is of the order of 30 K, which falls within the range of the measured activation energies for resistivity. As one approaches the metal-insulator crossover, the potential barrier separating neighboring droplets (see Eq.~\eqref{eq:xi-loc}) decreases. Holes near the Fermi level in the potential wells can be thermally excited above the potential barrier to energies above the percolation threshold; this is an alternate mechanism for transport as against the usual interdroplet tunneling followed by Coulomb blockade.
Fig.~\ref{fig:barrier} shows the dependence of the interdroplet potential barrier on the dimensionless parameter $R_c/R_p.$ The dependence is approximately linear in $R_c,$ similar to the findings in Ref.~\onlinecite{efros93}.

\begin{figure}
\includegraphics[width=0.95\columnwidth]{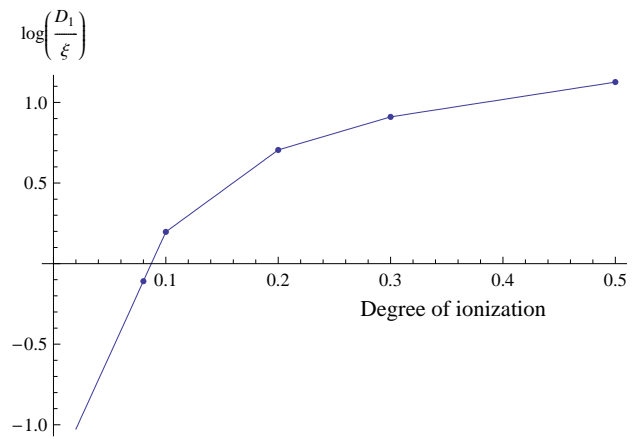}
\caption{\label{fig:met-ins}Plot of the calculated value of $\log(D_{1}/\xi)$ against the degree of ionization for Sample 3 at 77 K. The metal-insulator transition corresponds to $\log(D_{1}/\xi)=0.$}
\end{figure}

We end this Section with a few more words on the effect of partial ionization of the dopants. Using the reduced dopant density $n_{a}'$ does not affect the droplet sizes significantly but it does reduce $R_{c}$ by a factor $\sqrt{n_{a}'/n_{a}},$ thus bringing the system closer to the metallic percolation transition. Therefore, even small variations of $n_{a}'$ can strongly affect the potential barrier separating the droplets.
In Fig.~\ref{fig:met-ins}, we show  the interdroplet separation (in units of the localization length) as a function of the degree of ionization for Sample 3 at 77 K. The metal-insulator transition occurs when the effective degree of ionization is in the range from 0.05 to 0.1.

Since we are looking at the charge distribution at the moment, we
have ignored magnetism. Magnetism, and its effect on transport will
be considered in the next section.

\section{Resistivity}\label{sec:resistivity}

The droplet picture developed in the previous section can be used to understand the experimentally observed temperature dependence of the resistivity of the insulating samples. We will specifically study the resistivity of Samples 1 and 4 where the holes are well-localized in a droplet phase. In addition to localization effects, we will also need to take into account the effect of ferromagnetic correlations.

To our understanding, temperature dependent transport in insulating 2D DMS heterostructures has not yet been theoretically studied although numerous studies of corresponding nonmagnetic heterostructures exist in the literature. In the absence of magnetism, as in many disordered insulators, the temperature dependence of resistivity is expected to be of variable-range hopping type at very low temperatures and of Arrhenius type at higher temperatures. In the Arrhenius regime, one could either have nearest-neighbor tunneling together with an activation energy of the order of the mean droplet level spacing, $\Delta,$ or the classical thermal excitation over the barrier (see Fig.~\ref{fig:barrier}) separating neighboring droplets. Our resistivity measurements will not distinguish the two mechanisms and we will henceforth denote the Arrhenius energy fitting the data by $E_A;$ and $E_A = \Delta$ for the tunneling mechanism, which was estimated in the previous section. Next, we analyze the behavior of resistivity across the Curie temperature $T_{C}$, below which the ferromagnetic correlations increase rapidly. There is no continuous transition to a ferromagnetic state in two dimensions at a finite temperature and $T_{C}$ is a characteristic energy scale of the order of the exchange interaction associated with the ferromagnetism. Since $T_{C}$ is in the vicinity of 30 K, we are in the Arrhenius regime. This was experimentally observed for Samples 1 and 4.

In the absence of magnetism, the resistivity would behave as $\rho(T)\sim e^{E_A/T}$. When the droplets are magnetically polarized, there is an additional energy cost associated with introducing an extra charge into a given droplet if the spin orientation of the electrons in the droplet differs from that of the extra charge. Suppose the droplets are individually polarized (with different orientations) and let $\theta_{ij}$ be the angle between the magnetizations in the droplets at sites $i$ and $j.$ When a hole tunnels between these two droplets, the extra energy cost $\Delta_{ij}^{\text{mag}}$ at the destination droplet depends on the relative orientations of the magnetizations
\begin{align}
\Delta_{ij}^{\text{mag}} & =J(1-\cos\theta_{ij}).\label{eq:delta-mag}\end{align}
If the magnetic order in the droplet is induced by the Mn layer,
then $J$ is related to local magnetization in the Mn layer. If the
magnetic order is mainly determined by interaction of holes in the
quantum well, then $J$ is related to the exchange interaction in
the droplets. Our analysis is not dependent on the mechanism of ferromagnetism since $J$ and $T_{C}$ are phenomenological parameters. The temperature dependence of resistivity is governed by the total energy $E_{A}+\Delta_{ij}^{\text{mag}}$ associated with introducing an extra charge carrier into the droplet $j$ from a neighboring droplet $i$
\begin{align}
\rho(T) & \approx Ae^{E_{A}/T+J(1-\langle\cos\theta_{ij}\rangle)/T},
\label{eq:rho-T1}\end{align}
where we have approximated $\langle e^{-\cos\theta_{ij}/T}\rangle\approx e^{-\langle\cos\theta_{ij}\rangle/T}.$
For a two-dimensional ferromagnet, $\langle\cos\theta_{ij}\rangle=e^{-D_{1}/\xi_{M}(T)},$
where~\cite{arovas88,takahashi90,yablonskiy91}
\begin{align}
\xi_{M}(T) & =\left\{ \begin{array}{l}
a/\sqrt{1-T_{C}/T},\qquad T \gg T_{C}\\
a\exp[\pi T_{C}/2T],\qquad T \ll T_{C}\end{array}.\right.\label{eq:mag-correl}\end{align}
Here $a\sim 1/\sqrt{n_d}$ is a length scale of the order of interatomic separation
of the Mn dopants for the first ferromagnetic mechanism and interdroplet distance for the second mechanism and $T_{C}$ is the Curie temperature, below which the ferromagnetic correlations increase rapidly. If ferromagnetism is intrinsic to the Mn layer, then because of disorder we expect the local ferromagnetic interaction $J$ to be larger than the global transition temperature $T_{C}.$ For a homogeneous distribution of Mn atoms, $J\sim T_{C}$ for the same mechanism of ferromagnetism. If ferromagnetism is due to indirect exchange mediated by the holes, then $T_{C}$ falls with interdroplet tunneling probability and is smaller than $J$ in general.
\begin{figure}
\includegraphics[width=0.95\columnwidth]{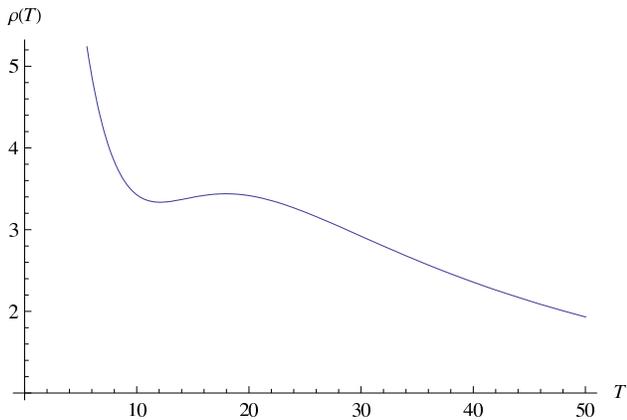}
\caption{\label{fig:illustration-hump} Calculated resistivity (in arbitrary units) as a function of temperature. Parameters from Sample 1 were used. We assumed a degree of ionization of $0.1,$ ferromagnetic transition temperature $T_{C}=30$ K, exchange integral $J=70$K. The peak in the resistivity occurs at a temperature lower than $T_{C}.$}
\end{figure}

\begin{figure}
\includegraphics[width=0.95\columnwidth]{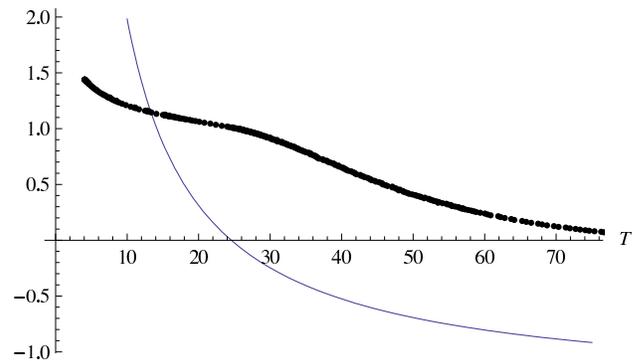}
\caption{\label{fig:corrlen} Plot of $\log(\xi_{M}/D_{1})$ showing the variation of magnetic correlation length $\xi_{M}$ for Sample 1 as a function of temperature (solid line). The dots show $\log[\rho(T)/\rho(77K)]$ for the same sample. The anomaly in the resistivity $\rho(T)$ occurs in the vicinity of the temperature where $\xi_{M}/D_{1}\sim 1$.}
\end{figure}

\begin{figure}
\includegraphics[width=0.95\columnwidth]{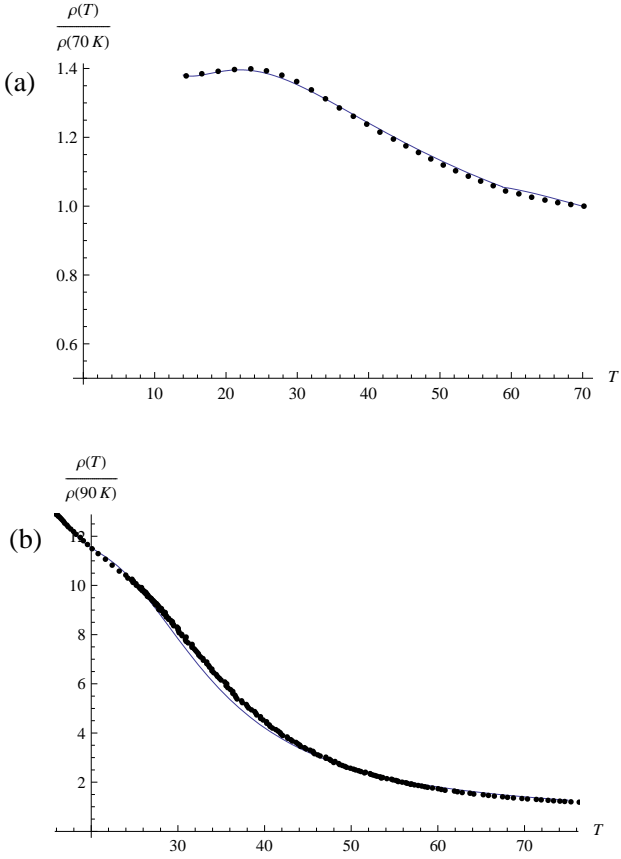}
\caption{\label{fig:fits}Observed temperature dependence of resistance for (a) Sample 4, in units of the resistance at $70$ K, and (b) Sample 1, in units of the resistance at $90$ K (points), and theoretical fits (solid lines). Sample 4 is near the percolation threshold and Sample 1 is well-insulating. The fits were made using Eq.~\eqref{eq:rho-T1}. Parameters such as the activation energy $E_{A}$ and the droplet separation $D_{1}$ were chosen close to the values obtained from the droplet model. The magnetic parameters $J$ and $T_{C}$ were then varied to obtain the above fits. In both cases, the best fit value of $T_{C}$ was significantly larger than the temperature, at which the resistance anomaly (hump or shoulder) was observed. At lower temperatures, the resistivity becomes variable-range hopping type (not taken into account in our model). For Sample 4 in panel (a), the values used for the fit are  $D_{1}=2$  nm, $E_{A}=9$ K, $J=39$ K, and $T_{C}=30$ K; for Sample 1 in panel (b), the parameters are  $D_{1}=9.4$ nm, $E_{A}=51$ K, $J=56$  K, and $T_{C}=49$ K.}
\end{figure}

Fig.~\ref{fig:illustration-hump} represents the calculated resistivity as a function of temperature. While we assumed a $T_{C}$ of 30 K, the peak in the resistivity appears at a significantly lower temperature. This is a characteristic feature of the 2D DMS heterostructures in contrast with bulk DMS where the peak appears near $T_{C}$ (and usually at temperatures higher than $T_{C}$). In the bulk case, while the peak does not also coincide with $T_{C},$ it is nevertheless possible to obtain the $T_{C}$ based on resistivity measurements~\cite{jungwirth06}. The key physical difference is that the magnetic correlation length for bulk DMS diverges upon approaching $T_{C}$ from higher temperatures, whereas in 2D, the magnetic correlation length remains finite except at $T=0$ (see Eq.~\eqref{eq:mag-correl}) due to the absence of true long-range ferromagnetic order at finite temperatures in 2D. The peak is related to the temperature when the magnetic correlation length
becomes comparable to the interdroplet separation: the temperature
corresponding to this peak is determined by the specific values of
parameters of the sample and not only by $T_{C}.$ In Fig.~\ref{fig:fits}, we show the observed resistance and a fit based on our model for Samples 1 and 4. Fig.~\ref{fig:corrlen} shows that the shoulder in the resistivity of Sample 1 occurs near the temperature where $\xi_{M}/D_{1}\sim 1$.

While making the fits, we made a number of observations. The position of the peak or shoulder is sensitive to $T_{C}$ and $D_{1},$ while
$J$ and $E_{A}$ determine the sharpness of the resistance anomaly.
If $D_{1}$ is reduced, then $\langle\cos\theta_{ij}\rangle=e^{-D_{1}/\xi_{M}(T)}$
changes from $0$ to $1$ at a higher temperature, which shifts the
resistance anomaly to a higher temperature. However, reducing $D_{1}$ also decreases the sharpness of the anomaly since then $e^{-D_{1}/\xi_{M}(T)}$ changes much more slowly with temperature. While this can be addressed to some extent by increasing $J,$ that in turn corresponds to a much larger (Arrhenius) activation energy at higher temperatures. Increasing $T_{C}$ also shifts the anomaly to a higher temperature but this happens without making the anomaly less sharp or increasing the Arrhenius energy. We found that values of $D_{1}$ and $E_{A}$ chosen near the calculated values generally gave good fits.

We conclude this Section with some observations about the metallic Samples 2 and 3 in the context of existing work on bulk metallic DMS systems. Temperature-dependent resistivity of bulk, metallic DMS samples in the vicinity of the Anderson transition has been analyzed in Ref.~\onlinecite{moca10}. Such an approach may be generalized to our two-dimensional heterostructures. While in 2D we do not have an Anderson transition, scaling theory for 2D may be used instead, and ferromagnetism effects can be incorporated as in the 3D calculation. In essence, we make use of the 2D scaling equation for the conductance,
\begin{align}
g(\xi) & =g(\xi_{0})-\frac{e^{2}}{\pi^{2}\hbar}\ln(\xi/\xi_{0}),
\label{eq:scaling}\end{align}
where $\xi$ is the localization length and $\xi_{0}$ is a short
length scale corresponding to $g(\xi_{0})$, together with the relationship
\begin{align}
g(\xi) & =e^{2}\nu(\epsilon_{F}){\cal D}\approx e^{2}\nu(\epsilon_{F})(k_{B}T)\xi^{2},\label{eq:hopping}\end{align}
which describes inelastic hopping transport in a regime where the
Thouless energy ${\cal D}/\xi^{2}\sim k_{B}T.$ The dependence on
magnetism is through $g(\xi_{0},T,H)\approx g_{0}(1+qM^{2}(T/T_{C},H))$ as in Ref.~\onlinecite{moca10}, where $M$ is the magnetization.

\section{Discussion and conclusions}\label{discussion}

We studied the effect of disorder, Coulomb interaction and ferromagnetism on the transport properties of 2D heterostructures $\delta$-doped by Mn.

The observation of Shubnikov--de Haas oscillations for fields perpendicular to the 2D direction of the quantum well confirmed the two-dimensionality of our samples. Resistivity measurements as well as previously measured magnetic hysteresis~\cite{aronzon07,aronzon08-2,aronzon08-3} and anomalous Hall effect~\cite{aronzon10,aronzon08} established magnetic ordering at low temperatures.  Our samples spanned the percolative metal-insulator transition region ranging from the very insulating to metallic behavior. We put the main emphasis on the two most insulating samples. The insulating samples are particularly interesting to us for they provide valuable insights into the mechanism of ferromagnetism in the DMS heterostructures.

We showed how at low carrier density the interplay of disorder in the spatial distribution of the dopant atoms and screening effects by the holes in the 2D quantum well leads to electronic phase separation in the quantum well. For this phase, we obtained the typical size of the hole droplets, their mean separation, and their energy levels.
Unlike conventional nonmagnetic GaAs/AlGaAs heterostructures, a two-subband model was used here as the carrier density was much larger in these heterostructures. We introduced a simple nearest-neighbor hopping model for the resistivity of this droplet phase taking into account the discreteness of the energy levels in the droplets and the effect of ferromagnetic correlations between spins on neighboring droplets. The values of the parameters in the resistivity model were obtained from droplet model where possible.
The ferromagnetic parameters such as the Curie temperature were varied to fit the observed data. A good agreement with the experiments was obtained. To our understanding, ours is the first theoretical study of the transport properties of 2D DMS heterostructures.

An important understanding that emerged from our study concerns the
relation between the position of the peak or shoulder in the resistivity data and the Curie temperature. Unlike 3D DMS systems where such resistance features are found in the vicinity of the Curie temperature (and above it), we showed that in 2D DMS heterostructures, the peak or shoulder-like features are not completely determined by the Curie temperature, and furthermore, the Curie temperature is typically substantially larger than the temperature at which such features are observed. Physically,
this is because the resistivity changes once the magnetic correlation length becomes comparable to the interdroplet separation. Clearly,
the divergence between the Curie temperature and the position of the
resistivity peak will be stronger for the more insulating samples.

Our calculations of the resistivity are independent of the microscopic mechanism of ferromagnetism since $J$ and $T_{C}$ are phenomenological parameters. Nevertheless a study of the dependence of $J$ and $T_{C}$ on sample parameters such as $p,$ $n_{a},$ $\lambda$ can give us crucial clues. Two main possibilities are that (a) the ferromagnetic ordering takes place in the Mn layer by some intrinsic mechanism such as the Zener indirect exchange mediated by holes in the $\delta$-layer~\cite{menshov09}, whereas the holes in the 2D transport layer merely respond to the Mn spin polarization,
and (b) the ferromagnetic ordering of Mn atoms is mediated significantly by the holes in the 2D transport layer~\cite{meilikhov08}. For mechanism (a), the transition temperature will be insensitive to the spacer thickness $\lambda$ as well as $p.$ However, one must be very careful in obtaining the transition temperature information from the resistance data. The position of the anomaly in resistance, sensitive to both $D_{1},$ and $T_{C},$ and $D_{1}$ is affected by the spacer thickness $\lambda.$ Thus, even if $T_{C}$ is independent of $\lambda,$ the position of the anomaly in the resistance data does depend on $\lambda$. A better test of mechanisms (a) and (b) is possible with the insulating samples. Note that as the sample becomes more insulating, the separation $D_{1}$ of the droplets will increase and the interdroplet tunneling will decrease exponentially with $D_{1}.$ If the magnetism is mediated by the holes in the 2D layer, then $T_{C},$ which is of the order of the strength of the magnetic interaction will be proportional to the interdroplet tunneling probability, which depends on the carrier density as $e^{-C/\sqrt{p}}$ since $D_{1}+2R_{p1}\propto1/\sqrt{p}$. Another clue is provided by the values of $T_{C}$, which we extracted from the fits of the resistance data for Samples 1 and 4 with our model. We found that $T_{C}$ for the more insulating Sample 1 was significantly larger than that of Sample 4. One way to understand this result is by observing that the Mn doping density $n_{a}$ is larger in Sample 1, which implies stronger magnetic interaction of the Mn atoms. This seems to support the possibility (a) of an intrinsic mechanism. ``Metallic'' samples discussed here are more likely to have both mechanisms contributing to ferromagnetism. We have seen that the localization length $\xi_z$ for the hole wavefunction in the GaAs region is of the order of 1 nm in all our samples. This compares well with the estimated localization length in Ref.~\onlinecite{meilikhov08}. An RKKY mechanism of ferromagnetism in a metallic sample leads to a $T_C$ for Mn atoms in the $\delta$-layer that decreases exponentially with the distance to the spacer layer as $T_C(\lambda)\approx T_C(0)e^{-4\lambda/\xi_z}$.  Mn atoms lying closer to the quantum well can however give larger Curie temperatures.

We thus believe that the insulating samples are ideally suited for
resolving the question of mechanism of ferromagnetism. In addition
to resistivity measurements, temperature dependent anomalous Hall
effect measurements can give us valuable clues to the mechanism
of ferromagnetism. It would be very interesting to compare the values of the Curie temperature extracted from our resistivity fits and from the temperature dependence of the anomalous Hall effect.

\section*{Acknowledgments}

We are grateful to E.Z. Meilikhov and V.V. Tugushev for helpful discussions.

The work was supported by the Russian Foundation for Basic Research, projects 09-02-92675-IND and No. 11-02-00363 and by the Indo-Russian Collaboration Program, grant No. INT/RFBR/P-49. We also acknowledge the partial support of EuroMagNET under EU Contract No. RII3-CT-2004-506239. V.T. and K.D. acknowledge the support of TIFR.

\end{document}